# MatlabCompat.jl: helping Julia understand Your Matlab/Octave Code

Vardan Andriasyan[1], Yauhen Yakimovich[1], Artur Yakimovich[1*]

Affiliations:

[1]University of Zurich

*correspondence: artur.yakimovich@uzh.ch, a.yakimovich@ucl.ac.uk

Scientific legacy code in MATLAB/Octave not compatible with modernization of research workflows is vastly abundant throughout academic community. Performance of non-vectorized code written in MATLAB/Octave represents a major burden. A new programming language for technical computing Julia, promises to address these issues. Although Julia syntax is similar to MATLAB/Octave, porting code to Julia may be cumbersome for researchers. Here we present MatlabCompat.jl - a library aimed at simplifying the conversion of your MATLAB/Octave code to Julia. We show using a simplistic image analysis use case that MATLAB/Octave code can be easily ported to high performant Julia using MatlabCompat.jl.

## Motivation and significance

### Scientific background and the motivation for developing the software

MATLAB/Octave is a fourth-generation multi-paradigm numerical programming language. MATLAB™ is a computing environment and a proprietary programming language (MathWorks Inc.). Designed initially mostly for matrix manipulations, its current features allow data visualization, algorithms implementation, user interface creation, and interfacing program languages. Introduced in the end of 1980s its syntaxial sibling, GNU Octave, offered an open source alternative to the commercial product of Mathworks Inc. (Eaton, Bateman, and Hauberg 1997). More than a tool for engineers, MATLAB/Octave has become a *de facto* default tool for developing scientific code, often used to solve problems that are otherwise typical for general purpose programming languages. This lead to an accumulation of prototype-like scientific software solutions unable to scale up to the promise and, thus, limiting the research work. To address this, in recent years a number of new programming languages and frameworks have been proposed including SciPy for Python (Jones et al. 2001) (Olivier, Rohwer, and Hofmeyr 2002), SCILAB (Campbell, Chancelier, and Nikoukhah 2009), FreeMat, jLab, Rlab and others. While some projects managed to successfully undertake the effort of rewriting their research software into one of these new frameworks (Carpenter et al. 2006) (Kamentsky et al. 2011), for many this represents a major problem due to programming language differences. A newly introduced open source dynamic high level technical programing language Julia aims at shattering the paradigm that high level prototype code has to be inherently inefficient (Bezanson et al. 2012) (Bezanson et al. 2014). Arguably, unlike MATLAB/Octave, it is by-design effective for general-purpose programming (either as a specification language or utilized as a web language).

Julia aims at fulfilling the requirements of a language for high-performance numerical and scientific computing. To this end, Julia allows concurrent, parallel and distributed computing. It uses eager evaluation, is garbage-collected, and includes efficient libraries for linear algebra, regular expression matching, floating-point calculations, fast Fourier transforms, and random number generation. Beyond the rich basic package Julia has currently 770 packages registered, which span functionalities from audio signal processing, through CUDA interface, to Bioinformatics. Despite the young age Julia has been featured in several research papers in fields ranging from Computer Science and Technical Computing (Heitzinger and Tulzer 2014) (Knopp 2014) (Foulds et al. 2013) to Biochemistry (Baldassi et al. 2014). Having a type system with parametric types, being a dynamic programming language, and using multiple dispatch as its core programming paradigm are distinctive features of Julia language design. It provides simple ways to call C and Fortran libraries directly and, unlike MATLAB/Octave, without unnecessary glue code. Reported in (Bezanson et al. 2012) in 2012 Julia had attracted 550 mailing list subscribers, 1500 GitHub followers, 190 forks, and more than 50 total contributors. To date the number of followers increased to 6251, forks to 1396 and contributors to 428.

Unlike other high level technical computing languages Julia offers symbolic meta-programming (similar to LISP or Haskell), rich datatype system with templates support, multiple dispatch of methods allowing reusing them dynamically, and other attractive features of a modern programming language. The source code written in Julia is executed using just-in-time (JIT) compilation based on low level virtual machine (LLVM) (Lattner and Adve 2004) bytecode, which may deliver performance matching performance of the iconic C language (Bezanson et al. 2012) (Bezanson et al. 2014).

Much like MATLAB/Octave, types in Julia are run time inferred, the algorithm used for such dynamic type inference and implementation details are described in (Kaplan and Ullman 1980) (Bezanson et al. 2012). Authors mention, types may be used to make declarations, but their declaration is not required for performance (Bezanson et al. 2012). This is achieved in several dynamic optimization passes addressing the method input/output typing and targeted LLVM code generation. The latter includes standard scalar optimizations, such as strength reduction, dead code elimination, jump threading, and constant folding (see (Bezanson et al. 2012) for the details). These points may provide an incentive for the developers to switch from MATLAB/Octave to Julia. Similar to *Esperanto* of human languages, the syntax of Julia combines syntaxes of MATLAB/Octave, Python, R, Perl and other languages. However, the code written in MATLAB/Octave requires significant modifications to be understood by Julia.

## Software importance for solving scientific problems and contribution to the process of scientific discovery

Technical and scientific computing serves as the most fundamental scientific tool for tasks ranging from quantification of experimental results, to simulation of scientific hypotheses and making predictions for scientific fields ranging from Biology to Particle Physics. A large fraction of scientific questions in these fields are addressed using MATLAB/Octave language.

Here we present MatlabCompat.jl a Julia language library aimed at simplifying the conversion of MATLAB/Octave based research software to the novel Julia technical computing language by providing one-way compatibility and MATLAB/Octave to Julia code translating modules. MATLAB/Octave is widely used to develop complex research software, e.g. in the field of image analysis (Kiss et al. 2014) (Carpenter et al. 2006). A very specific example of a scientific question to be addressed is the automated quantification of micrographs of cultured cells obtained in a phenotypical high-throughput screening (HTS). Due to a very large number of images (Rämö et al. 2014), it is impossible to perform such quantifications manually. Quantification of HTS images is addressed using among others MATLAB/Octave (Yakimovich et al. 2012) (Carpenter et al. 2006) (Rämö et al. 2014).

Using MatlabCompat.jl this scientific question can now be addressed potentially more efficiently with Julia. Beyond image analysis MatlabCompat.jl spans a handful of functions including input/output, working with strings, linear algebra and others providing researchers with an easy-to-reach MATLAB/Octave runtime alternative. MatlabCompat.jl allows researchers to easily convert their MATLAB/Octave code into Julia and enrich their code using Julia language. The added values of such code conversion efforts include using features like dynamic types, dataframes, as well as, allowing for performance gain. To illustrate how MatlabCompat.jl helps solving typical scientific problem of experimental results quantification, below we provide a typical use case from image analysis of micrographs obtained with an automated epifluorescent screening microscope (based on published data (Yakimovich et al. 2012)).

## Related work

One or two-way compatibility between different programming languages can be achieved either by manual or automatic translation, application program interface (API) based integration or, in cases of exceptional syntax similarity, by compatibility libraries. Multiple efforts on automatic translation of programming languages date to as early as 1960s (Ledley and Wilson 1962) (Irons 1961). API based integrations exists between and for multiple programming languages and platforms, including e.g. Python (Autin et al. 2012), MATLAB (Bornstein et al. 2008). Julia languages has multiple APIs based integrations working and in development, including MATLAB.jl allowing calling MATLAB™ using MATLAB™ engine from Julia environment. Finally, compatibility libraries typically written in the destination language are a very common way to translate not just code but also principles and practices well established in one language to another language, like e.g. MATLAB™ plotting library for Python (Hunter 2007) (Barrett et al. 2005). While calling MATLAB/Octave code from Julia is made possible by the MATLAB.jl package, there was no tool thus far assisting in conversion of MATLAB/Octave code to Julia language. Automated program translation for MATLAB/Octave into other high level programming languages poses additional practical challenges. Namely the absence of the official parsing grammar, runtime-dependent grammar parsing (interpretation), absence of strict datatyping and others.

As suggested in (Moynihan and Wallis 1991), a comprehensive high-level to high-level language translation is generally unrealistic, but should the effort be undertaken capabilities of such a converter must be chosen carefully by weighing implementation costs. In the specific case of MATLAB/Octave to Julia translation we aim for a pragmatic one-way transliteration following 'transliterate and refine' strategy described in (Waters 1988). We argue that the latter makes sense in the specific case of MATLAB/Octave to Julia translation due to the likeness of the syntax. The practical challenge of such a translator lays in ensuring compatibility across the toolboxes (more than 60 in case of MATLAB) code bases of MATLAB/Octave and Julia. To this end, the rapidly growing Julia community (discussed below and observable on GitHub) has or is actively implementing overlapping functionalities independently via more than 770 Julia libraries (e.g. Images.jl etc.). Syntax wrapping hence can be an easy way to achieve toolbox compatibility (in this paper illustrated e.g. by functions compatible with "Image Processing Toolbox" by Mathworks Inc.). We argue that this means that the compatibility across the entire code base in the particular case of MATLAB/Octave to Julia translation may realistically be achieved as a community effort.

We propose to address the MATLAB/Octave to Julia compatibility using MatlabCompat.jl library package. We call for collaborative assistance on growing the functionality of the package, which can be accomplished by forking our source code through Github and pull-requesting your improvements later on (see details on contributing below). Additionally, MatlabCompat.jl is set up as an organization on Github aimed at uniting multiple researcher working on the package and creating a developers community.

## Software description

MatlabCompat.jl is a Julia language package that helps converting your code in Matlab/Octave into Julia. Much like Julia language itself MatlabCompat.jl is fully Github based and a work in progress inviting everyone to collaborate. Following a Github workflow it is set up as a Github organization, which is open for new members and the library source code is open for Github based pull requests. The longterm goal of the library is to allow effortless compatibility of Matlab/Octave research code to Julia.

A full (semantic) code conversion for high level programming languages is a highly non-trivial task. A more pragmatic and, thus, successful strategy would involve either partial translation/adaptation of the syntax of the source language to the syntax of the destination (object) language (Ledley and Wilson 1962) or usage of the library substituting (translating) the code of the source language "as is" in the environment of the destination language (**Fig. 1**). While first strategy is always available for a developer it's possibilities can be quickly exhausted. Therefore, the rest of the compatibility effort in MatlabCompat.jl is covered by the second strategies.

More often than not, certain code adaptation may become necessary or desired for several reasons. First, both MatlabCompat.jl and Julia language itself are in an early stage. Therefore, MatlabCompat.jl might either not cover the necessary functionality of Matlab/Octave or this functionality might be non-compatible for syntaxial and other reasons. Second, pursuing exclusively strategy 2 is very challenging in development. Third, pursuing exclusively strategy 2 developers risk creating non-optimal *"stub"* Julia code.

MatlabCompat.jl is helpful to simplify refactoring effort in transition from legacy Matlab/Octave code to new Julia code. The compatibility library aims to minimize naming conflicts between Matlab/Octave code reserved namespace and respective taken identifiers in the namespace of Julia. We refer to such property in newly converted code (strategy 1, see **Fig. 1**) as minimal target-invasive (Julia-invasive) compatibility. Only in newly written or refactored (strategy 2, see **Fig. 1**) application code developers get fully exposed to utilization of all Julia features. Therefore developers are suggested to pursue hybrid strategy requiring code refactoring to make their Matlab/Octave work in Julia yet minimizing the effort needed to achieve it.

## Software Architecture

Following the minimal Julia invasivity concept mentioned, MatlabCompat.jl has a well-structured architecture allowing minimization of namespace conflicts with Julia's *Base* package (**Fig. 2**). A typical usage of a Julia package involves invoking *"using"* command:

```
using MatlabCompat
```

The namespace conflicts between Julia and Matlab/Octave are partially inevitable given the syntaxial closeness of Julia and Matlab/Octave. E.g. the function *max()* in Matlab/Octave is closer to function *maximum()* in Julia, while *max()* function name is taken in Julia. To address this type of issues Julia language avoids global declarations and uses local scopes. Typically for other Julia packages invoking using with a package would export all the functions of the package into the scope. Yet in case of MatlabCompat.jl to avoid namespace conflicts with *"Base"* package, invoking the line above for MatlabCompat package would export into the scope only those functions names of which are not conflicting with *"Base"* package. To substitute the respective function from the *"Base"* package by a function from MatlabCompat.jl package with the same name (for Matlab/Octave compatibility reasons) a user would have to explicitly *import* the respective module of MatlabCompat.jl with the same naming prefix of this module or a particular symbol within such module into the scope of the converted program. Nonetheless, most of these special cases are explicitly handled by symbol export within MatlabCompat.jl. For example to use max() function from MatlabCompat.jl package rather than from *"Base"* package - write:

```
using MatlabCompat, MatlabCompat.MathTools
```

```
max([1 2 3])
```

MatlabCompat.jl follows a rather simple modules organization convention. Namely, all the modules are named with function purposes based self-explanatory names in *"CamelCase"* with capitalized first letter. Modules may contain submodules if necessary. All functions modules contain should be either aliases (wrappers) or re-implementations of Matlab/Octave functions named the same way. The only module representing an exception in this case is module *"Support"*, which contains helper-functions, not having an analogue neither in Matlab/Octave nor in Julia languages, necessary for other functions in the package.

## Software Functionalities

Currently the major functionalities of the MatlabCompat.jl include syntax aliases or reimplementations of Matlab/Octave functions organized in modules with self-explanatory naming: *"StringTools"*, *"MathTools"*, *"ImageTools"*, *"Support"* (see **Fig. 2** and help). Like the Julia language, MatlabCompat.jl is in an active development and is actively receiving new contributions, including new modules. For an up-to-date overview of functions provided by different modules of the package please see the help. For pragmatic reasons the Matlab/Octave to Julia compatibility functions covered by MatlabCompat.jl at the current stage are strongly shifted towards image analysis and processing functions, including e.g. *graythresh()* for Otsu thresholding (Otsu 1975), *bwlabel()* for labeling and subsequent counting of e.g. bright objects over dark background (discussed in detail below) and other functions.

## Sample code snippets analysis

An example of functionality provided by the MatlabCompat.jl package standing out, but worth mentioning is covered by function *rossetta()* (*"Support" module*). This function is aimed to be used as the starting point for Matlab/Octave-to-Julia conversion of your code. It is aimed to parse m-files (first argument) making basic syntax changes required for the code compatibility to Julia and then save the jl-file (second argument) to the specified path:

```
using MatlabCompat

rossetta(janus.m, janus.jl)
```

One can test the functionality of this code using the minimalistic janus.jl program discussed below. Unless currently in development, all the functions provided by the MatlabCompat.jl package have respective test programs that exercise all the main features of the package for automated testing. Functionality of the whole package is continuously automatically tested using Travis Continuouse Integrations (CI) service.

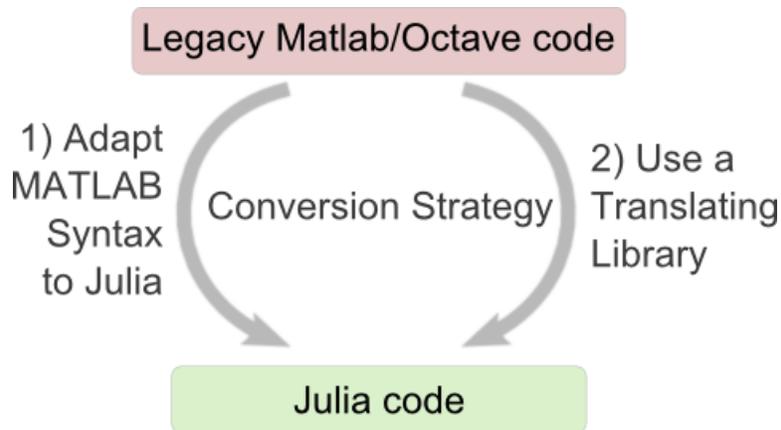

*Figure 1. MATLAB to Julia Code Conversion Strategies*

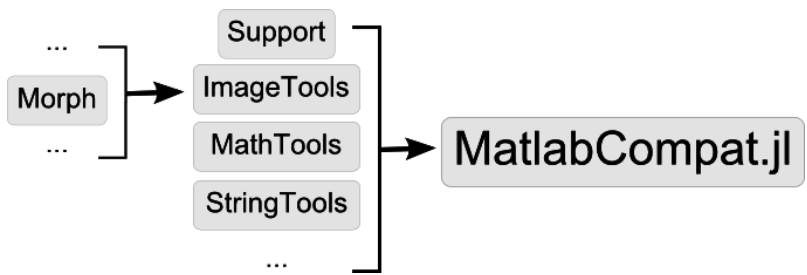

*Figure 2. MatlabCompat Library Modules Inclusion Model Allows minimizing Namespace Conflicts*

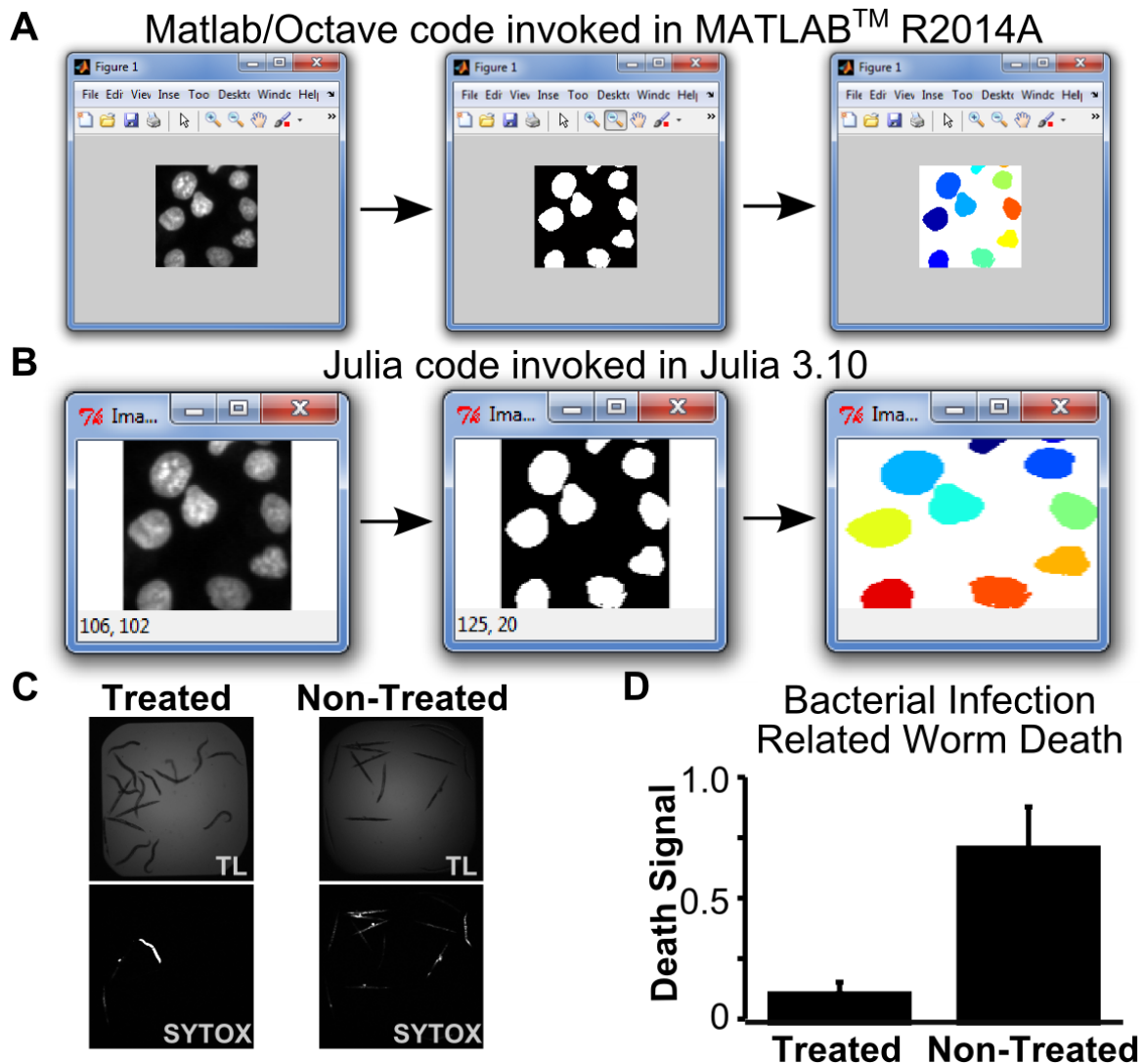

*Figure 3. Step-by-step Output of the Model Use Case Program "janus".* *(A) - Output in MATLAB[TM] R2014A. (B) - Output in Julia 0.3.10. (C) - Illustrative example of how simple code in MATLAB/Octave and it's translated version in Julia can be used to perform automatic quantification of a dataset published in (Moy et al. 2009), where the authors performed an in vivo high-throughput screen for conditions preventing bacterial infection related death of Caenorhabditis elegans model organisms. Based on their published dataset, worms placed in 384 microtiter plates, infected with equal amounts of Enterococcus faecalis. Respective wells were treated with either ampicillin (treated, positive control for infection related death inhibition) or DMSO (non-treated, mock control) or a compound of unknown efficacy (not shown here). Each well was imaged in transmission light (TL, unspecific wide-field image) or*

*SYTOX Orange fluorescence (specific for dead worms) (Moy et al. 2009), representative micrographs of which are shown. (D) - Here we have quantified from the published raw micrographs of SYTOX Orange fluorescence micrographs for either treated or non-treated control wells using either "janus2.m" or automatically translated "janus2.jl" microprograms with identical results. Death signal on the graph is measured as number of SYTOX Orange positive pixels min/max normalized between 0 and 7000. Error bars correspond to the standard deviation from three different experimental replica wells micrographs for each condition measured in the published dataset (Moy et al. 2009).*

## Illustrative Examples

To demonstrate the functionalities of the MatlabCompat.jl library we have created *janus.m* - a minimalistic image analysis program in Matlab/Octave (see **supplementary video S1**). First, the program is reading a remote image - a micrograph of cultured cells, stained with a nuclear stain (A549 human lung carcinoma cells stained with Hoechst 33342 dye from a dataset used it (Yakimovich et al. 2012)). Next the program is using Otsu (Otsu 1975) thresholding to segment (separate) bright foreground pixels from dark background pixels. Finally, the program counts connected components in the foreground pixels, which in an idealistic case is equal to the number of cellular nuclei in the picture. **Fig. 3A** shows MATLAB™ R2014a output upon invoking *janus.m*. Code listing for *janus.m* together with commentaries (separated by the symbol "%" in Matlab/Octave) is provided here. To ensure that the same program can be understood by Julia using MatlabCompat.jl package either a parsing step or minimal changes to the code are necessary. These changes include basic syntax adaptation like substitution of single quotes symbol (') by double quotes (")

```
rosetta(janus.m, janus.jl)
```

In case of *janus.m* the function *"rosetta()"* generates immediately Julia interpretable code (see **Fig. 3B** for output in Julia 0.3.10) provided here.

To show how similarly simplistic microprogram, which can be readily automatically translated from MATLAB/Octave to Julia , can be used to address a research question in a different scientific field we have used another published dataset of an image-based screen performed for roundworm *Caenorhabditis elegans* model organism (Moy et al. 2009). The authors of that screen aimed to look for novel small compounds inhibiting *Enterococcus faecalis* bacterial infection. Their readout was fluorescence signal of SYTOX Orange dye staining specifically dead worms, while living worms remained impermeable to that dye - thus emitting little or no fluorescent signal. Death of the experimental *C. elegans* worms in that study was attributed to *E. faecalis* bacterial infection (**Fig. 3C**). For illustration purposes, of how a simple MATLAB/Octave or Julia programs can measure such a readout we have first created the *janus2.m* MATLAB/Octave program counting foreground pixels from a fluorescence micrograph (**Fig. 3D**), code for which is provided here. Next, similarly to the previous example, we invoked *"rosetta()"* function on the *janus2.m* to automatically translate it to Julia language, yielding *janus2.jl* provided here.

Both programs gave identical fluorescence micrographs quantification results shown in **Fig. 3D**, suggesting that, as expected by the design of this published study (Moy et al. 2009), the worm death was lower in the case of treatment with the control inhibitor of *E. faecalis* bacterial infection compared to the mock control. Altogether, these examples suggest, that in the specific case of MATLAB/Octave to Julia conversion, results, readily applicable to complex research questions, can be achieved rather quickly. While, as discussed before, this might not work equally well in every case, this conversion may give better results with further development of the MatlabCompat.jl by its community.

Arguably, one of the most attractive features Julia may provide for the research software is the potential for performance optimization. To challenge the results published on the official website of the language we have executed the published standard algorithms performance tests in slightly different environment for a new version of Julia (0.3.10) and two different versions of MATLAB™ (R2014a and R2015a). The tests include Fibonacci recursion (fib), parse integers (parse_int), Mandelbrot set: complex arithmetic and comprehensions (mandel), numeric vector sorting (quicksort), slow $\pi$-series (pi_sum), random matrix statistics (rand_mat_stat), large random number generation and multiplication of random matrices (rand_mat_mul). Results (**supplementary text S2**) suggest small differences in execution time. The differences are most pronounced for Julia's most efficient Fibonacci Recursion (fib). This can be attributed to the difference in test settings and the version of the language. Despite the small differences, we were able to faithfully reproduce the pattern and the orders of magnitudes of the published results. Surprisingly MATLAB™ R2015a performed slightly slower than R2014a in our test environment. We have next compared the execution times of our model image analysis program identical to *"janus.m"* and *"janus.jl"*. Median results of 1000 iterations in the test environment mentioned above suggested that MATLAB™ R2014a performed 0.29 folds of Julia 0.3.10 and R2015a 0.30 folds of Julia 0.3.10. We concluded that performance of Julia and MATLAB™ was comparable for this specific case. This may be caused by the fact that MATLAB™ is highly optimized for vector operations used in this program.

## Impact

Research code written in Matlab/Octave is used in many scientific and engineering disciplines. Multiple scripts written and used in research groups across the world serve in research tasks from data acquisition to analysis and visualization. Porting the Matlab/Octave code to Julia using [MatlabCompat.jl](MatlabCompat.jl) may significantly broaden the applicability of their code. Features of Julia language can significantly improve research process and the exchange of tools and datasets. These features include flexible Perl-like regular expression for text parsing, R-like dataframes for working with large scientific, Python-like (IPython) browser based integrated development environment (IJulia package), ability to turn prototype code into a web-applications, easy-to-use web based cloud computing environment provided by [Juliabox](Juliabox) and many others. Performance optimization of the code combined with using the cloud solutions for Julia for a non-computational research group can mean budget and infrastructure needs could differ between few laptops and a full scale computer cluster needed to accomplish a research project. This, in turn, can have profound implications for feasibility of a large number of research projects.

MatlabCompat.jl has been included in Julia's official packages and is widely discussed across the community. We hope attract more code developers to the crowd-based MatlabCompat.jl development. For this purpose the development of the package is based on Github organization rather than an individual account. By this we hope to establish a community board and maximize the involvement into the project.

## Conclusions

We have developed MatlabCompat.jl - a Julia package aimed to assist conversion of Matlab/Octave code to Julia providing one-way compatibility between these languages. Being a library package, MatlabCompat.jl code is by definition reusable and ment to be taken up by the community. We welcome contributions from the community following the contribution guide. Quality of our software is continuously monitored by test programs that exercise all the main features of the software in a CI environment. Furthermore, all the code written using MatlabCompat.jl can easily be portable to other platforms including mobile through JuliaBox web service. We show practical applicability of the software on a simplistic Matlab/Octave image analysis script ("janus.m"). This scripts is readily executable in MATLAB™ R2014a and, upon simple parsing by a MatlabCompat.jl function (*rosetta()*) converting it to "janus.jl", the script is readily executable in Julia 0.3.10.

We were able to reproduce performance comparison tests published on the official website of Julia within the order of magnitude of execution time, suggesting a much higher performance of Julia implementation of Fibonacci recursion over MATLAB™. While our test environment was different, we argue that it was closer to the "out-of-the-box" user experience on a modern desktop computer. Furthermore, we believe it was important to reconfirm the performance test results in a slightly different (potentially less optimized) environment to show the stability and reproducibility of those results.

The three orders of magnitude faster performance of Julia compared to MATLAB™ can be explained by the fact that Fibonacci recursion is a great representation of algorithms programmed using nested loops. However, performance of our test use case program was slightly lower in Julia, than in MATLAB™. This can be explained by the fact that MATLAB™ is highly optimized for vectorized tasks. These test results suggest a word of caution: not every type of code converted from Matlab/Octave will be faster in Julia compared to MATLAB[TM]. However, this might change once it will be made possible to easily generate binaries from the source code - i.e. static compilation.

Altogether, our results suggest that converting Matlab/Octave code to Julia may significantly improve performance of research software requiring large amount of iterations. However, this is a subject to algorithms used, software implementation, input data and other details. MatlabCompat.jl can assist in accomplishing such a conversion effort. We invite the community to help further develop MatlabCompat.jl, which can ultimately improve the functionality and performance of the research software.

## Acknowledgements

We would like to thank Prof. Urs Greber and Dr. Ivo Sbalzarini for scientific discussions and motivation for completing the work. Additionally we would like to thank Julia community for help, recognition and the warm welcome.

# Required Metadata

## Current code version

Required metadata listing can be found in **Tab. 2**

*Code metadata*

| Nr. | Code metadata description | Please fill in this column |
| --- | --- | --- |
| C1 | Current code version | v0.1.2 |
| C2 | Permanent link to code/repository | Source code |
| C3 | Legal Code License | MIT |
| C4 | Code versioning system used | git (Github) |
| C5 | Software code languages, tools, | Julia |
| C6 | Compilation requirements | NA |
| C7 | Link to developer documentation/manual | Manual |
| C8 | Support email for questions | artur.yakimovich@uzh.ch |

# References

Eaton, John Wesley, David Bateman, and Søren Hauberg. 1997. Gnu Octave. Network thoery.

Jones, Eric, Travis Oliphant, P Peterson, and others. 2001. "Open Source Scientific Tools for Python." Scipy.

Olivier, B.G., J.M. Rohwer, and J.-H.S. Hofmeyr. 2002. Molecular Biology Reports 29 (1/2). Springer Science $\mathplus$ Business Media: 249–54. doi:10.1023/a:1020346417223.

Campbell, Stephen L., Jean-Philippe Chancelier, and Ramine Nikoukhah. 2009. "Modeling and Simulation in Scilab." In Modeling and Simulation in Scilab/Scicos with ScicosLab 4.4, 73–106. Springer New York. doi:10.1007/978-1-4419-5527-2_3.

Carpenter, AE, TR Jones, MR Lamprecht, C Clarke, IH Kang, O Friman, DA Guertin, et al. 2006. "CellProfiler: Image Analysis Software for Identifying and Quantifying Cell Phenotypes.." Genome Biol 7: R100.

Kamentsky, L, TR Jones, A Fraser, MA Bray, DJ Logan, KL Madden, V Ljosa, C Rueden, KW Eliceiri, and AE Carpenter. 2011. "Improved Structure, Function and Compatibility for CellProfiler: Modular High-Throughput Image Analysis Software.." Bioinformatics 27: 1179–80.

Bezanson, Jeff, Stefan Karpinski, Viral B Shah, and Alan Edelman. 2012. "Julia: A Fast Dynamic Language for Technical Computing." ArXiv Preprint ArXiv:1209.5145.

Bezanson, Jeff, Alan Edelman, Stefan Karpinski, and Viral B Shah. 2014. "Julia: A Fresh Approach to Numerical Computing." ArXiv Preprint ArXiv:1411.1607.

Heitzinger, Clemens, and Gerhard Tulzer. 2014. "Julia and the Numerical Homogenization of PDEs." In 2014 First Workshop for High Performance Technical Computing in Dynamic Languages. Institute of Electrical & Electronics Engineers (IEEE). doi:10.1109/hptcdl.2014.8.

Knopp, Tobias. 2014. "Experimental Multi-Threading Support for the Julia Programming Language." In 2014 First Workshop for High Performance Technical Computing in Dynamic Languages. Institute of Electrical & Electronics Engineers (IEEE). doi:10.1109/hptcdl.2014.11.

Foulds, James, Levi Boyles, Christopher DuBois, Padhraic Smyth, and Max Welling. 2013. "Stochastic Collapsed Variational Bayesian Inference for Latent Dirichlet Allocation." In Proceedings of the 19th ACM SIGKDD International Conference on Knowledge Discovery and Data Mining - KDD 13. Association for Computing Machinery (ACM). doi:10.1145/2487575.2487697.

Baldassi, Carlo, Marco Zamparo, Christoph Feinauer, Andrea Procaccini, Riccardo Zecchina, Martin Weigt, and Andrea Pagnani. 2014. "Fast and Accurate Multivariate Gaussian Modeling of Protein Families: Predicting Residue Contacts and Protein-Interaction Partners." Edited by Kay Hamacher. PLoS ONE 9 (3). Public Library of Science (PLoS): e92721. doi:10.1371/journal.pone.0092721.

Lattner, C., and V. Adve. 2004. "LLVM: A Compilation Framework for Lifelong Program Analysis & Transformation." In International Symposium on Code Generation and Optimization 2004. CGO 2004. Institute of Electrical & Electronics Engineers (IEEE). doi:10.1109/cgo.2004.1281665.

Kaplan, Marc A., and Jeffrey D. Ullman. 1980. "A Scheme for the Automatic Inference of Variable Types." Journal of the ACM 27 (1). Association for Computing Machinery (ACM): 128–45. doi:10.1145/322169.322181.

Kiss, Alexa, Peter Horvath, Andrea Rothballer, Ulrike Kutay, and Gabor Csucs. 2014. "Nuclear Motility in Glioma Cells Reveals a Cell-Line Dependent Role of Various Cytoskeletal Components." Edited by Christof Markus Aegerter. PLoS ONE 9 (4). Public Library of Science (PLoS): e93431. doi:10.1371/journal.pone.0093431.

Rämö, Pauli, Anna Drewek, Cécile Arrieumerlou, Niko Beerenwinkel, Houchaima Ben-Tekaya, Bettina Cardel, Alain Casanova, et al. 2014. "Simultaneous Analysis of Large-Scale

RNAi Screens for Pathogen Entry." BMC Genomics 15 (1). Springer Science $\mathplus$ Business Media: 1162. doi:10.1186/1471-2164-15-1162.

Yakimovich, A, H Gumpert, CJ Burckhardt, VA Lütschg, A Jurgeit, IF Sbalzarini, and UF Greber. 2012. "Cell-Free Transmission of Human Adenovirus by Passive Mass Transfer in Cell Culture Simulated in a Computer Model.." J Virol 86: 10123–37.

Ledley, Robert S., and James B. Wilson. 1962. "Automatic-Programming-Language Translation through Syntactical Analysis." Commun. ACM 5 (3). Association for Computing Machinery (ACM): 145–55. doi:10.1145/366862.366872.

Irons, Edgar T. 1961. "A Syntax-Directed Compiler for ALGOL 60." Commun. ACM.

Autin, Ludovic, Garth Johnson, Johan Hake, Arne Olson, and Michel Sanner. 2012. "UPy: a Ubiquitous CG Python API with Biological-Modeling Applications." Computer Graphics and Applications, IEEE 32 (5). IEEE: 50–61.

Bornstein, B. J., S. M. Keating, A. Jouraku, and M. Hucka. 2008. "LibSBML: an API Library for SBML." Bioinformatics 24 (6). Oxford University Press (OUP): 880–81. doi:10.1093/bioinformatics/btn051.

Hunter, John D. 2007. "Matplotlib: A 2D Graphics Environment." Computing in Science & Engineering 9 (3). Institute of Electrical & Electronics Engineers (IEEE): 90–95. doi:10.1109/mcse.2007.55.

Barrett, Paul, John Hunter, J Todd Miller, J-C Hsu, and Perry Greenfield. 2005. "Matplotlib–A Portable Python Plotting Package." In Astronomical Data Analysis Software and Systems XIV, 347:91.

Moynihan, Vincent D., and Peter J. L. Wallis. 1991. "The Design and Implementation of a High-Level Language Converter." Softw: Pract. Exper. 21 (4). Wiley-Blackwell: 391–400. doi:10.1002/spe.4380210405.

Waters, R.C. 1988. "Program Translation via Abstraction and Reimplementation." IEEE Transactions on Software Engineering 14 (8). Institute of Electrical & Electronics Engineers (IEEE): 1207–28. doi:10.1109/32.7629.

Otsu, Nobuyuki. 1975. "A Threshold Selection Method from Gray-Level Histograms." Automatica 11 (285-296): 23–27.

Moy, Terence I., Annie L. Conery, Jonah Larkins-Ford, Gang Wu, Ralph Mazitschek, Gabriele Casadei, Kim Lewis, Anne E. Carpenter, and Frederick M. Ausubel. 2009. "High-Throughput Screen for Novel Antimicrobials Using a Whole Animal Infection Model." ACS Chem. Biol. 4 (7). American Chemical Society (ACS): 527–33. doi:10.1021/cb900084v.